\documentclass[11pt,a4paper,reqno]{amsart}
\usepackage{amssymb,amsmath,amsthm}
\usepackage{graphicx}
\usepackage{verbatim}
\usepackage{url}
\usepackage[usenames,dvipsnames]{color}
\usepackage{amsmath}
\usepackage{empheq}
\numberwithin{equation}{section}
\newtheorem{thm}{Theorem}    
\theoremstyle{definition}
\newtheorem{examp}[thm]{Example}
\theoremstyle{definition}

\newcommand{\R}{\mathbb{R}}

\newcommand{\cE}{\mathbb{E}}
\numberwithin{equation}{section}
\begin{document}
\title{Calculating Variable Annuity Liability `Greeks' Using Monte Carlo Simulation}  
\author{\large Mark J. Cathcart$^{\dag}$, Steven Morrison$^{\ddag}$ and Alexander J. M\lowercase{c}Neil$^{\dag}$\\
\\
$^{\dag}$ Heriot-Watt University and the Maxwell Institute for Mathematical Sciences. $^{\ddag}$ Barrie and Hibbert Limited.}  
\date{\today}  
\begin{abstract}
Hedging methods to mitigate the exposure of variable annuity products to market risks require the calculation of market risk sensitivities (or ``Greeks''). The complex, path-dependent nature of these products means these sensitivities typically must be estimated by Monte Carlo simulation. Standard market practice is to measure such sensitivities using a ``bump and revalue'' method. As well as requiring multiple valuations, such approaches can be unreliable for higher order Greeks, e.g., gamma. In this article we investigate alternative estimators implemented within an advanced economic scenario generator model, incorporating stochastic interest-rates and stochastic equity volatility. The estimators can also be easily generalized to work with the addition of equity jumps in this model.
\end{abstract}
\maketitle
\pagestyle {plain}
\section{Introduction}
Many issuers of variable annuity (VA) contracts employ a hedging strategy to mitigate some of the risk borne out of writing these products. The challenge of estimating the sensitivities of VA liabilities to their key risk drivers is, therefore, an important issue to many practitioners. For the purposes of this article a VA contract is ``any unit-linked or managed
fund vehicle which offers optional guarantee benefits as a choice for the
customer'' \cite{VApaper}. The complex nature of the guaranteed benefits included in these products means that numerical techniques must be used to calculate such sensitivities. Monte Carlo simulation is perhaps the most flexible of these approaches.

In this article we extend the standard approaches to estimating option price sensitivities by Monte Carlo simulation to the problem of estimating the sensitivities of the liability of a stylized VA product. Furthermore, we adapt these approaches to be compatible with a more realistic economic model than the Black-Scholes framework, incorporating stochastic volatility and interest-rates. The main contributions of this article are as follows. The likelihood ratio method, a standard approach for evaluating option price sensitivities by Monte Carlo simulation, is extended to work under our more realistic market model with stochastic volatility \emph{and} stochastic interest-rates. This builds on work by Broadie and Kaya \cite{BroadKaya}, who only consider stochastic volatility. We also provide a full derivation of the equations for the likelihood ratio weights, which cannot otherwise be found in the literature. Also, the pathwise method is developed in the context of VA liabilities, following a similar approach to Hobbs et.\thinspace al.\thinspace \cite{Hobbs}, but considering a more complex VA product and the more realistic economic model. Finally, the pathwise and the likelihood ratio approaches are then combined together to construct a new and efficient mixed estimator for the second-order gamma sensitivity of the VA liability. This will be important to practitioners looking to hedge their VA exposures, as obtaining an accurate and unbiased gamma estimate will allow the insurer to adapt their hedging strategy portfolio to take account of the convexity of their liabilities with respect to key market exposures. 

The paper is organized as follows; In Section \ref{Background} a overview of the standard methods for estimating option price sensitivities by Monte Carlo simulation is given. Three standard approaches are reviewed: the bump and revalue method, which is the natural finite difference approach often used in practice; the pathwise method, which was developed in the context of option pricing by Brodie and Glasserman \cite{BrGla}; the likelihood ratio method, developed in the context of option pricing by Broadie and Glasserman \cite{BrGla} and Glasserman and Zhao \cite{GlaZh}. Mixed hybrid estimators, introduced by Broadie and Glasserman \cite{BrGla}, which combine the latter two of these standard approaches together to construct an efficient estimator for second-order sensitivities, will also be reviewed. 

In Section \ref{sect:CLRE} the likelihood ratio method is extended to the setting of a more sophisticated economic model. This is an original contribution and builds on the work of Broadie and Kaya  \cite{BroadKaya}. In Section \ref{VALiabSens} the standard approaches of Section \ref{Background} and the extension of the likelihood ratio method in Section \ref{sect:CLRE} are developed for VA liability sensitivities. The pathwise approach for VA liabilities follows a similar approach to the article of Hobbs et.\thinspace al.\thinspace \cite{Hobbs}, except that the present article considers a more complex product and the stochastic model for volatility and interest-rates.  The likelihood ratio method is then extended to the liability of our stylized VA product. Next, the mixed estimators are constructed for the VA liability gamma sensitivity.

To conclude the article, Section  \ref{VAtests} compares all the estimators developed for the stylized VA product in Section \ref{VALiabSens} in terms of numerical efficiency. The mixed gamma sensitivity estimator is found to be particularly efficient, which is appealing as this is the sensitivity for which the standard approach performs the poorest.  

\section{ Option Price Sensitivity Estimators}\label{Background}
The liability from a VA contract
takes a form very similar to the payoff of an option (or the sum of
the payoffs of a series of options). Thus, in attempting to develop
efficient methods for determining these VA liability
sensitivities it would seem sensible to look for guidance from Monte
Carlo methods for estimating option price sensitivities
which have been developed by other researchers in recent years. We now summarize three classes of approach for calculating
the sensitivities of options by simulation. Firstly, we introduce a simple
method commonly adopted by insurers to estimate VA liability Greeks; the ``bump \& revalue'' approach.

\subsection{``Bump \& Revalue'' Approach}
This term refers to the concept of simulating the cost of the option under some base scenarios for key risk-drivers and then again under `bumped scenarios', i.e., with the sensitivity parameter increased by some small perturbation, say $\Delta \theta$. This sensitivity parameter could be the initial equity index level for a VA product, say. This is just a forward difference estimate in the sensitivity parameter. If the function $Y(\theta)$ gives the discounted payoff of an option, then the price of this option is then given by $\alpha(\theta) = \cE[Y(\theta)]$ with respect to a pricing measure. Now let $Y_1(\theta), \ldots, Y_n(\theta)$ represent the discounted option payoff along simulation paths $1,\ldots,n$, the estimator of the first-order sensitivity would then be given by
\begin{equation*}
\hat{\Delta}^{\small\mbox{B}} = \frac{\bar{Y}(\theta+\Delta \theta) - \bar{Y}(\theta)}{\Delta \theta}
\end{equation*}
for the chosen `bump size' $\Delta \theta$, where $\bar{Y}(\theta)$ is the average of $Y_1(\theta),\ldots,Y_n(\theta)$. The expectation of this estimator is given by
\begin{equation*}
\cE\big[\hat{\Delta}^{\small\mbox{B}}\big] = \frac{\alpha(\theta+\Delta \theta) - \alpha(\theta)}{\Delta \theta}.
\end{equation*}

The variance will be significantly reduced if the same random number stream is used in calculating the `bumped' and `base' option prices, as oppose to using independent random number streams. The `bump and revalue' approach can incur fairly large sampling errors, particularly with second-order sensitivity estimates.  Therefore, alternative approaches for estimating sensitivities are required by practitioners. Glasserman \cite{Glass} introduces two more sophisticated general approaches to determining the sensitivities of option prices. 

\subsection{Pathwise Estimator}\label{sectpathwise}
The first of these approaches is the pathwise derivative method and the estimator of the first-order sensitivity with respect to $\theta$ is given as follows. We can find the derivative of $\alpha(\theta)=\cE[Y(\theta)]$ analytically along each simulation path using
\begin{equation}\label{PWest}
Y'(\theta) = \lim_{h\to 0}\frac{Y(\theta + h) - Y(\theta)}{h}.
\end{equation}
If the interchanging of differentiation and taking expectations is justified, that is if
\begin{equation*}
\cE\bigg[\frac{d}{d\theta}Y(\theta)\bigg] = \frac{d}{d\theta}\cE[Y(\theta)]
\end{equation*}
then $\frac{1}{n}\sum_{i=1}^{n}Y_i'(\theta)$ is an unbiased estimator of $\alpha'(\theta)$. 
\begin{examp}
As an example of this method, consider the delta of a call option under the Black-Scholes model (i.e., the sensitivity of the option price with respect to $S_0$ the initial underlying asset value). This can, of course, be found analytically, but helps illustrate the method. The discounted payoff of a call option is given by
\begin{equation*}
Y = e^{-rT}\max(S_T-K,0)
\end{equation*}
where $r$, $K$, $T$ and $S_T$ are the risk-free rate, strike price, time to maturity and equity price at maturity, respectively. Under Geometric Brownian Motion
\begin{equation}\label{GBM}
S_T=S_0e^{(r-\frac{1}{2}\sigma^2)T+\sigma\sqrt{T}Z}
\end{equation}
where $Z$ is a standard normal variate. Applying the chain rule to differentiate $Y$ with respect to $S_0$ with all other parameters held fixed gives
\begin{equation}
\frac{dY}{dS_0} = \frac{dY}{dS_T}\frac{dS_T}{dS_0}.
\end{equation}

With the term $\frac{dY}{dS_T}$ the derivative fails to exist at the strike price, however the event $\{S_T=K\}$ has measure zero and hence $\frac{dY}{dS_T}=e^{-rT}I(S_T>K)$ almost surely, where $I(A)$ represents the indicator function of event $A$. For the second-term, Equation \ref{GBM} gives $\frac{dS_T}{dS_0}=\frac{S_T}{S_0}$. Thus, the pathwise estimator for the Black-Scholes delta is
\begin{equation*}
\frac{dY}{dS_0} = e^{-rT}I(S_T>K)\frac{S_T}{S_0}.
\end{equation*}

If we wish to find the Black-Scholes gamma, i.e., sensitivity of the delta to the initial asset price, we have to differentiate $W = e^{-rT}I(S_T>K)$ with respect to $S_0$, which is equivalent to estimating the delta of a digital option. Here $W$ is differentiable with respect to $S_0$ with probability one, and takes the value zero. However, in this case
\begin{equation*}
0 = \cE\bigg[\frac{dW}{dS_0}\bigg] \neq \frac{d}{dS_0}\cE[W]
\end{equation*}
and the pathwise estimator is inapplicable (and is
generally inapplicable when the payoff is discontinuous or in estimating second-order derivatives). The change
in $\cE[W]$ with a change in $S_0$ is explained by the fact that it
could cause $S_T$ to become `in-the-money'. Glasserman \cite{Glass}
gives some technical conditions for when the pathwise method can
be applied, however a less rigorous `rule of thumb' is that it can
be used when the payoff is continuous in the parameter of interest.
\end{examp}

\subsection{Likelihood Ratio Estimator}\label{sectLRMintro}
The second more sophisticated approach to estimating option
sensitivities introduced by Glasserman \cite{Glass} is known as the likelihood ratio method (LRM).
The LRM approach relies on differentiating probability densities rather
than payoff functions. Thus it does not require smoothness in the
payoff function, as was required in the pathwise
derivative method. 

Suppose we have a discounted payoff $Y$ expressed
as a function $f(X)$, where $X$ is a
$m$-dimensional vector of different asset prices (or alternatively,
one asset price at multiple valuation dates).
Then, assuming that $X$ has a probability density $g$
with parameter $\theta$, taking
the expected discounted payoff with respect to this density gives
\begin{equation*}
\cE[Y] = \cE[f(X(\theta))] = \int_{\R^m}f(x)g_{\theta}(x)dx.
\end{equation*}
Now, similar to the pathwise derivative approach, we assume the
order of differentiation and integration can be interchanged. Here,
however, this is not such a strong assumption, as typically densities
are smooth functions, whereas payoff functions are not. This gives
\begin{eqnarray*}
\alpha'(\theta) = \frac{d}{d\theta}\cE[Y] &=& \int_{\R^m}f(x)\frac{d}{d\theta}g_{\theta}(x)dx\\
                                 &=& \int_{\R^m}f(x)\frac{\frac{d}{d\theta}g_{\theta}(x)}{g_{\theta}(x)}g_{\theta}(x)dx\\
                                 &=& \cE\bigg[f(X)\frac{\frac{d}{d\theta}g_{\theta}(X)}{g_{\theta}(X)}\bigg]\\
                                 &=& \cE\bigg[f(X)\frac{d}{d\theta}\ln(g_{\theta}(X))\bigg].
\end{eqnarray*}
Then, $f(X)\frac{d}{d\theta}\ln(g_{\theta}(X))$ gives the likelihood ratio estimator for the sensitivity with respect to the parameter $\theta$, and this estimator is unbiased. The term $\frac{d}{d\theta}\ln(g_{\theta}(x))$ is often known in statistics as the ``score function'' and in this context is often referred to as the likelihood ratio weight, since it multiplies the discounted payoff function to give the sensitivity estimator.

The likelihood ratio method will still be applicable and robust in
the case of options with discontinuous payoff functions (and
estimating second-order sensitivities) as this approach looks to
differentiate the probability density, rather than the payoff
function.  

\begin{examp}
To show how this approach works in practice let us again consider the problem of estimating the Black-Scholes delta. In this case the discounted payoff function is given by $f(S_T) = e^{-rT}\max(S_t-K,0)$ and the log-normal density function used to calculate $S_T$ is given by
\begin{equation*}
g(x) = \frac{1}{x \sigma \sqrt{T}} \phi\bigg( \frac{\ln\big(\frac{x}{S_0}\big)-(r-\frac{1}{2}\sigma^2)T}{\sigma \sqrt{T}} \bigg)
\end{equation*}
where $\phi(\cdot)$ represents the standard normal density function. In this case the score function for the parameter is:
\begin{equation}
\frac{d}{dS_0}\ln(g(x)) = \frac{\ln\big(\frac{x}{S_0}\big)-(r-\frac{1}{2}\sigma^2)T}{S_0\sigma^2 T}.\label{eqBSscore}
\end{equation}
Evaluating this function at $S_T$ and multiplying by the discounted payoff of the option gives the unbiased estimator of the Black-Scholes delta as:
\begin{equation*}
e^{-rT}\max(S_T-K,0)\frac{\ln\big(\frac{S_T}{S_0}\big)-(r-\frac{1}{2}\sigma^2)T}{S_0\sigma^2 T}.
\end{equation*}
As we simulate for $S_T$ using the relationship
\begin{equation*}
S_T = S_0e^{(r-\frac{1}{2}\sigma^2)T + \sigma \sqrt{T}Z}
\end{equation*}
this likelihood ratio estimator can be written
\begin{equation*}
e^{-rT}\max(S_T-K,0)\cdot\frac{Z}{S_0\sigma \sqrt{T}}
\end{equation*}
where the term $Z/(S_0\sigma\sqrt{T})$ is known as the LRM weight.
\end{examp}

One interesting point to note is the form of the payoff can be replaced by any other payoff function to give an estimator for the delta corresponding to the option with that payoff. For a digital option under the Black-Scholes model we just multiply the payoff function for this option with the LRM weight which we have already calculated, giving the estimator for delta as
\begin{equation*}
e^{-rT}I(S_T>K)\cdot\frac{Z}{S_0\sigma \sqrt{T}}.
\end{equation*}
This is a general feature of the likelihood ratio method which makes
it appealing as a technique for estimating option sensitivities with
different types of payoff in the same simulation run. Note that for reference, the LRM weight for the option gamma, found by an analogous process of differentiating the density $g_{\theta}(x)$ twice with respect to $\theta$, can be easily shown to be  $\frac{Z^2-Z\sigma\sqrt{T}-1}{S_0^2\sigma^2T}$.

For path-dependent options the LRM weights have a similar structure to the equivalent European option LRM weight for the required sensitivity, except they will only use information along the path up until the first valuation date. The following Example demonstrates this.

\begin{examp}
The analysis thus far has studied how the LRM could be developed for European-type payoffs, however many VA products on offer in the markets will exhibit some degree of path dependency in their liabilities.  With this in mind let us turn our attention to studying how the LRM extends to path-dependent options.

Following Glasserman \cite{Glass}, let us consider estimating the sensitivities of an arithmetic Asian call option with $m$ valuation dates. The payoff on this option at maturity is then given by 
\begin{equation*}
Y = f(S_1,\ldots,S_m) = e^{-rT}\max(\bar{S} - K,0), \hspace{0.15in}\mbox{where }\hspace{0.05in}\bar{S}=\frac{1}{m}\sum_{t=1}^{m} S_t.
\end{equation*}
Using the Markov property of Geometric Brownian motion, the underlying density for the equity asset path can be factorized as
\begin{equation*}
g(x_1,\ldots,x_m) = g_1(x_1|S_0) g_2(x_2|x_1) \cdots g_m(x_m|x_{m-1})
\end{equation*}
where $g_j(x_j|x_{j-1})$ is the transition density from time $t_{j-1}$ to $t_j$, i.e.,
\begin{eqnarray*}
g_j(x_j|x_{j-1}) &=& \frac{1}{x_j \sigma \sqrt{t_j - t_{j-1}}}\phi(\zeta_j(x_j|x_{j-1})) \\
\zeta_j(x_j|x_{j-1}) &=& \frac{\log(x_j/x_{j-1})-(r-\sigma^2/2)(t_j-t_{j-1})}{\sigma \sqrt{t_j - t_{j-1}}}.
\end{eqnarray*}

Suppose we wish to get an LRM estimator for the delta of the Asian option. From the above factorization it is clear that $S_0$ is a parameter of the first factor, $g_1(x_1|S_0)$, only. This means we can express the score function, corresponding to the delta sensitivity, as:
\begin{equation*}
\frac{\partial \log g(S(t_1),\ldots, S(t_m)}{\partial S_0} = \frac{\partial \log g_1(S(t_1)|S_0)}{\partial S_0} = \frac{\zeta_1(S(t_1)|S_0)}{S_0\sigma\sqrt{t_1}} = \frac{Z_1}{S_0\sigma\sqrt{t_1}}
\end{equation*}
where $Z_1=\zeta_1(S(t_1)|S_0)$ is the Gaussian increment which takes us from time zero to time $t_1$. Likewise, the LRM estimator of the gamma sensitivity has a score function component which only relies on the equity asset path out to the first valuation date.
\end{examp}

\subsection{Mixed Estimators for Second-order Sensitivities}
In estimating the second-order sensitivities of the option price with respect to some parameter, for example the gamma sensitivity, one can also combine the pathwise and LRM approaches to create hybrid mixed estimators.

\begin{examp}
Again let us consider estimating the gamma Greek of a European call option under the Black-Scholes model. Firstly, by applying the pathwise approach to the LRM delta estimator, we obtain
\begin{eqnarray*}
\Gamma^{\mbox{LR-PW}} &=& \frac{d}{dS_0}\bigg(e^{-rT}\max(S_T - K, 0)\frac{Z}{S_0\sigma\sqrt{T}}\bigg)\\
&=& e^{-rT}I(S_T > K)K\frac{Z}{S_0^2\sigma\sqrt{T}}.
\end{eqnarray*}

Alternatively applying the LRM to the pathwise delta estimator gives the PW-LR mixed gamma estimator. The pathwise delta estimator has both functional and distributional dependence on $S_0$. To capture the distributional dependence, the pathwise estimator is multiplied by the LRM weight. For the functional dependence another pathwise derivative is taken. This gives
\begin{eqnarray*}
\Gamma^{\mbox{PW-LR}} &=& e^{-rT}I(S_T > K)\frac{S_T}{S_0}\cdot\frac{Z}{S_0 \sigma\sqrt{T}} + \frac{d}{dS_0} \bigg(e^{-rT}I(S_T > K)\frac{S_T}{S_0}\bigg)\\
&=&e^{-rT}I(S_T > K)\frac{S_T}{S_0^2}\bigg(\frac{Z}{\sigma\sqrt{T}} - 1 \bigg).
\end{eqnarray*}
These mixed estimators typically give smaller standard errors for the gamma of European options, than the bump and revalue or pure LRM approaches \cite{Glass}.
\end{examp}

\section{Conditional Likelihood Ratio Estimator}\label{sect:CLRE}
In order to use the likelihood ratio method, or the mixed hybrid second-order sensitivity estimators, one must be able to determine the probability density
function of the underlying asset returns in order to derive the score function. With a Black-Scholes model, this
score function is easily found in closed-form, but it is well known that this model does not give a realistic description of true market dynamics (see, e.g., \cite{DuffPan}). There is, however, an elegant 
approach which allows us to incorporate stochastic volatility/interest-rates and jumps in the equity returns, whilst still being able to utilize the tractability of the Black-Scholes model.

The conditional likelihood ratio method (CLRM) was introduced by Broadie and Kaya \cite{BroadKaya} for the Heston stochastic volatility model \cite{Hest}. In this article we extend this further to also incorporate stochastic interest-rates through the Cox-Ingersoll-Ross (CIR) model. Stochastic jumps can also be added to the equity process, but have been omitted here for ease of illustration of the method. See Broadie and Kaya \cite{BroadKaya} for the structure of the likelihood ratio weights when jumps are present, but stochastic interest-rates are not. In full, the system of Stochastic Differential Equations which
govern our asset price dynamics is:
\begin{align}
dS_t &= r_t S_tdt + \sqrt{V_t}S_tdW_t^S\nonumber\\
dV_t &= \kappa_{V}(\theta_{V} - V_t)dt + \sigma_V\sqrt{V_t}dW_t^V\label{eqHestCIR}\\
dr_t &= \kappa_{r}(\theta_{r} - r_t)dt + \sigma_{r}\sqrt{r_t}dW_t^r.\nonumber
\end{align}
Thus, the variance and interest rate both follow the same stochastic
process (but calibrated with different parameters). The Brownian motions for the asset,
volatility and interest rate processes are correlated as
follows:
\begin{equation*}
\mbox{corr}(W_t^S,W_t^V)=\rho_{S,V}, \hspace{0.1in}\mbox{corr}(W_t^S,W_t^r)=\rho_{S,r} \hspace{0.05in}\mbox{ and }\hspace{0.05in}\mbox{corr}(W_t^V,W_t^r)=\rho_{V,r}.
\end{equation*}
In order to construct this correlation structure among the normally distributed
risk-drivers we employ a Cholesky decomposition. Let $(Z_1, Z_2, Z_3)$ be independent standard normal variates. Then standardized increments $(Z_V, Z_r, Z_S)$ for the variance, interest-rate and equity processes are constructed by setting 
\begin{equation}
(Z_v, Z_r, Z_S)' = A(Z_1, Z_2, Z_3).
\end{equation}
Here $A$ is the lower-triangular matrix satisfying $A A'= \rho$, where $\rho$ is the correlation matrix constructed from $\rho_{S,V}$, $\rho_{S,r}$ and $\rho_{V,r}$. The correlation matrix $\rho$ is assumed to be positive-definite. For completeness, the matrix $A$ is given by:
\begin{equation*}
A = \left( \begin{array}{ccc}
1 & 0 & 0 \\
\rho_{V,r} & \sqrt{1-\rho_{V,r}^2} & 0 \\
\rho_{S,V} & \frac{\rho_{S,r}-\rho_{S,V} \rho_{V,r}}{\sqrt{1-\rho_{V,r}^2}} & \sqrt{1-\rho_{S,V}^2-\frac{(\rho_{S,r}-\rho_{S,V} \rho_{V,r})^2}{1-\rho_{V,r}^2}} \end{array} \right)\
\end{equation*}

The random variable $Z_r$ will now have correlation $\rho_{V,r}$ with $Z_V$ and the random normal variate $Z_S$ will have
correlation $\rho_{S,V}$ with $Z_V$ and correlation $\rho_{S,r}$
with $Z_r$. This completes our Cholesky decomposition for the
correlation structure of the three risk-drivers out to the first time-step.
Now by conditioning on a realization of the variance and interest-rate processes, the asset returns become log-normal. One can then use the same form of LRM weights as for the Black-Scholes models, but with the addition of a couple of extra factors to account for the conditional information. Let us explore this idea further. Using the Cholesky
decomposition the asset price dynamics can now be expressed as:
\begin{eqnarray*}
\frac{dS_t}{S_t} &=& r_tdt + \sqrt{V_t}dW_t^S\\
     &=& r_tdt + \sqrt{V_t}\bigg( a_{3 1}dW_t^{I_1} + a_{3 2}dW_t^{I_2} + a_{3 3}dW_t^{I_3}\bigg)
\end{eqnarray*}
where $dW_t^{I_1}$, $dW_t^{I_2}$ and $dW_t^{I_3}$ are three independent Brownian motions, corresponding to $Z_{1}$, $Z_{2}$ and $Z_{3}$ in our discretization and $a_{i j}$ is the element in row $i$ and column $j$ of matrix $A$. This construction for $dS_t$ expresses the asset price
dynamics in terms of all the risk-drivers in the system of SDEs,
which together dictate the behaviour of the asset
returns. In theory an interest rate process with a larger
number of risk-factors could be used. One would employ a Cholesky
decomposition to correlate all the risk-drivers as required and
express the asset price dynamics, $dS_t$, in terms of all these
other risk-drivers. One would however do this using a
numerical Cholesky decomposition program to find the coefficients. 

To proceed towards a conditional log-normal representation of the asset returns, we need an expression for the stochastic process
representing the logarithm of the returns, $X_t$. This expression for $dX_t$, derived in Appendix A using It\^{o}'s
formula, is given as
\begin{equation*}
dX_t  = r_tdt + dY_t -\frac{V_t}{2}a_{33}^2dt + \sqrt{V_t}a_{33}\mbox{ $dW_t^{I_3}$}
\end{equation*}
with
\begin{equation*}
dY_t = -\frac{1}{2}V_t a_{31}^2dt -\frac{1}{2}V_t a_{32}^2dt
     + \sqrt{V_t}a_{31}dW_t^{I_1} + \sqrt{V_t} a_{32}dW_t^{I_2}.
\end{equation*}
Using this result for $dX_t$, an expression for $S_T$ in terms of the initial asset price $S_0$
can now be found as:
\begin{align*}
S_T = S_0 \exp(Y_T)\exp\bigg(\frac{1}{T}\int_0^Tr_tdt -\frac{1}{2}a_{33}^2\frac{1}{T}\int_0^T V_t dt + a_{33}\frac{1}{T}\int_0^T \sqrt{V_t}dW_t^{I_3}\bigg)
\end{align*} 
where $Y_T$ is given by
\begin{equation*}
-\frac{1}{2}a_{3 1}^2\frac{1}{T}\int_0^T V_tdt -\frac{1}{2}a_{3 2}^2\frac{1}{T}\int_0^T V_tdt + a_{3 1}\frac{1}{T}\int_0^T\sqrt{V_t}dW_t^{I_1} + a_{3 2}\frac{1}{T}\int_0^T\sqrt{V_t}dW_t^{I_2}.
\end{equation*}

By defining
\begin{eqnarray}
\xi_T &=& \exp{(Y_T)},\\\label{XiformulaIntOn}
\bar{\sigma}_T &=& \sqrt{\frac{a_{3 3}^2}{T}\int_0^T V_tdt},\\\label{SigmaformulaIntOn}
\bar{r}_T &=& \frac{1}{T}\int_0^Tr_tdt,\label{rformulaIntOn}
\end{eqnarray}
the price of a European call option under the Heston SV with CIR
interest rate model is given by
$\mbox{C}^{BS}(S_0\xi_T,K,\bar{\sigma}_T,\bar{r}_T,T)$, where $C^{BS}$ is the Black-Scholes analytical formula for the price of a European call option.
Also the CLRM weights can be determined
using these adjusted expressions for $S_0$, $\sigma$ and $r$ with
the Black-Scholes score function, for example Equation \ref{eqBSscore} for the delta sensitivity. This then gives us a means to obtain
Monte Carlo simulation based estimators of the sensitivities of
options under the proposed stochastic models for the underlying equity asset and
risk-free interest rate.

\begin{figure}[hbt]
\centering
\includegraphics[width=3in]{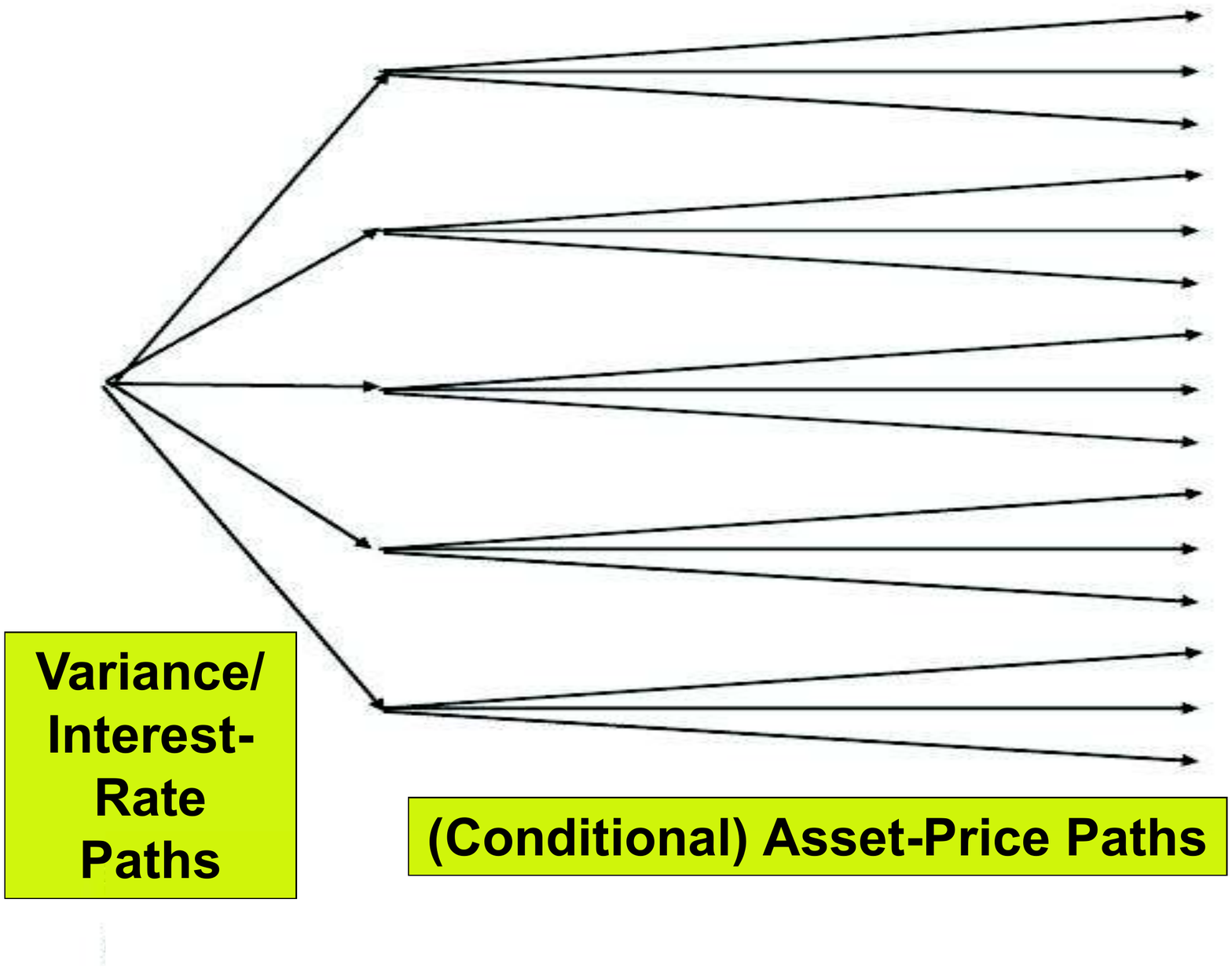}
\caption{}\label{NPgraph}
\end{figure}

Naturally, the simulation must be performed in a conditional structure, i.e., the expectation is taken as $\cE[\hat{\beta}] = \cE \Big[ \cE \Big[ \hat{\beta} \Big| v_1, \ldots, v_n, r_1, \ldots, r_n \Big] \Big]$. This set-up is illustrated in Figure \ref{NPgraph}.
\addtolength{\abovecaptionskip}{+5mm}
\addtolength{\belowcaptionskip}{+5mm}
\addtolength{\floatsep}{+5mm}

\section{Variable Annuity Liability Sensitivities}\label{VALiabSens}
\subsection{Example Variable Annuity Product}
The methods for estimating option price sensitivities will now be applied to the problem of estimating the sensitivity of the liabilities on a stylized variable annuity product. The idea behind this stylized example product is that it should be simple enough both in terms of tractability and ease of exposition, yet retain some of the key features which make these products so popular in many markets. It should also have liabilities which are path-dependent, as is typical of many VA products on the market. This example product, based on a product from \cite{VApaper}, will be of the Guaranteed Minimum Withdrawal Benefit (GMWB) type, where the policyholder is entitled to annual withdrawals from the underlying fund throughout the product lifetime, even if poor returns means that this fund diminishes to zero (at which point the insurance company must provide this income out its own reserves). 

More detail of this VA product and the policyholder will now be given. Firstly the contract
owner is a 65 year old male and from the start of the contract has
the `guarantee rider' activated, i.e., this GMWB option initially sits on the
pensions contract from the policyholder's 65th birthday. He will pay
an extra 1\% of the guarantee base (which is introduced below) in charges as a result of
this option being activated, on top of the annual fund management charge which is deducted from the VA fund. If the policyholder wishes to turn off
this option, he would be entitled to do so and this would cancel
this extra guarantee charge. We shall refer to this as a customer
`lapse'. The modelling of the rate of policyholder lapse is by no
means a trivial issue and is worthy of study in its own right. We will assume a constant level of policyholder lapse of 4\%, however a more realistic dynamic model of lapsation, based on fund levels and market conditions, could be employed without. This is a topic for further research.

Under the GMWB contract the policyholder is guaranteed to receive income at
the level of $w\%$ of the Guarantee Base each year after his 65th
birthday, until he turns off the guarantee `rider' or dies.
This Guarantee Base is initially set at the amount of the
policyholder premium, but can increase in value with an increasing VA
fund level for the first $\alpha$ years after annuitization. After this window
passes, the Guarantee Base will then remain at the same level for the
remainder of the product's lifetime. This ratchet feature, where the Guarantee Base steps up to the Fund Value if this is greater at the rebalancing date, is capped at a maximum year-on-year increase in the Guarantee Base of 15\%. This will become clearer when the cashflows are described mathematically shortly.

The underlying VA fund, which is initially funded by the
policyholder premium, is invested wholly in a single equity index with dynamics which are governed via the Heston-CIR model, given in Equations \ref{eqHestCIR}. Possible extensions of this analysis include investing in a mixture of equities and bonds, or in a portfolio of two different equity indices.

To model the cashflows on this policy mathematically let us define the Fund Value, Guarantee Base and policyholder income level at year $t$ after annuitization by $F_t$, $G_t$ and $I_t$ respectively. Also let $R_t$ denote the return from the equity index from year $t-1$ to year $t$ minus the management fees, i.e., $R_t = S_t/S_{t-1} -\eta$, where $\eta$ is the fund management charge, quoted as an annual percentage. Then we can track the level of Fund Value throughout the lifetime of the policy using the following equation:
\begin{equation}\label{VAcf1}
F_t = \max\big( (F_{t-1} - I_{t-1})(1+R_t),0\big).
\end{equation}
This expresses the Fund Value at year $t$ in terms of the Fund Value and Income level at year $t-1$. Naturally then, we must have starting Fund and Income levels to initiate this recursion. The initial Fund Value is just given by the policyholder premium at annuitization, $P$ (in units of the initial equity level), multiplied by the initial equity index level: $F_0 = P \cdot S_0 $.
The Guarantee Base at the end of year $t$ after annuitization can be expressed as
\begin{equation}\label{VAcf3}
G_t = I(t \leq \alpha )\min\bigg(\max(G_{t-1},F_t),1.15\times G_{t-1}\bigg) + I( t>\alpha)G_{t-1}.
\end{equation}
The income level the policyholder withdraws from the policy Fund Value at the end of year $t$ is then given by
$I_t = (w - \mu) G_t$.
Here, $w$ is a fixed parameter dictating the proportion of the Guarantee Base that is withdrawn by the policyholder at each annual re-balancing date and $\mu$ is the guarantee charge, taken as a percentage of the Guarantee Base each year. In our case we have $w = 4\%$ and $\mu = 1\%$.

The Liability, or Guarantee Shortfall, the insurer faces from issuing this VA contract on the market, measured at annuitization, can be expressed as
\begin{equation}\label{VAcf5}
L = \cE\Bigg[ \sum_{t=1}^{T} D_t p^{surv}_t \max(I_t - F_t,0) \Bigg]
\end{equation}
where $D_t$ is the factor to ensure the liability at each yearly withdrawal date is discounted back to annuitization, $p^{surv}_t$ is the probability of the policy remaining in force until year $t$ after annuitization (encompassing both the possibility of policyholder mortality and lapsation) and $T$ is the maximum contract term. Clearly, the insurer only faces a liability under the GMWB contract when the policyholder income cannot be met by the VA fund level. This is captured by the max function in the above formula for $L$, which sets the summand to zero at rebalancing/withdrawal dates where $F_t \geq I_t$.

\subsection{Pathwise VA Liability Estimator}\label{sectVApw}
We begin by developing a pathwise methodology for estimating the VA liability sensitivities, for the example VA product of the last section. This method, proposed for a simple VA product by Hobbs et.\thinspace al.\thinspace \cite{Hobbs}, is just the natural extension of the pathwise approach for option sensitivities to the case of a VA liability, $L$. The liability is analogous to a series of European options of increasing maturity, thus the same limitations of this approach for European option price sensitivities will apply here. Assuming that interchanging the order of differentiation and taking the expectation is justified the derivative of the liability with respect to the initial equity asset price can be expressed as 
\begin{equation}
\Delta_{\mbox{PW}} = \cE\Bigg[ \sum_{t=1}^{T} D_t p^{surv}_t \frac{\partial}{\partial S_0}\max(I_t - F_t,0). \Bigg]\label{VAcf6}
\end{equation}
The derivative only acts on the third term of the summand, and we can express the derivative of this term as
\begin{equation}
\frac{\partial}{\partial S_0}\max(I_t - F_t,0) = I(I_t > F_t)\cdot \Big( \frac{\partial I_t}{\partial S_0} - \frac{\partial F_t}{\partial S_0} \Big).\label{VAcf7}
\end{equation}
The problem of estimating the delta sensitivity of the VA liability is now one of estimating the derivative of the Fund Value, $F_t$, and income level, $I_t$, for each year, $t$, after annuitization. Appealing to the structure of the product's cashflows these derivatives must be calculated recursively. Using Equation \ref{VAcf1} we can express the derivative of the year $t$ Fund Value with respect to $S_0$ as
\begin{equation*}
\frac{\partial F_t}{\partial S_0} = \max\bigg(\Big( \frac{\partial F_{t-1}}{\partial S_0}-\frac{\partial I_{t-1}}{\partial S_0} \Big)(1 + R_t),0\bigg).
\end{equation*}
Similarly, using Equation \ref{VAcf3} the derivative of the year $t$ income level with respect to $S_0$ for the example product can be expressed using
\begin{eqnarray*}
\frac{\partial G_t}{\partial S_0} &=& I\big((F_t \geq G_{t-1})\cap (F_t\leq 1.15\times G_{t-1})\cap (t\leq \alpha)\big) \frac{\partial F_t}{\partial S_0}\\
& &  + I\big((F_t \geq G_{t-1})\cap (F_t>1.15\times G_{t-1})\cap (t \leq \alpha)\big) \times 1.15 \times \frac{\partial G_{t-1}}{\partial S_0}\\
& &  + I\big((F_t<G_{t-1})\cap (t \leq \alpha)\big) \frac{\partial G_{t-1}}{\partial S_0} + I(t \geq \alpha) \frac{\partial G_{t-1}}{\partial S_0}
\end{eqnarray*}
where $A \cap B$ is the intersection of events $A$ and $B$, and the derivative of the income level is then given by $\frac{\partial I_t}{\partial S_0} = (w-\mu) \cdot \frac{\partial G_t}{\partial S_0}$.

\vspace{0.15in}
These recursions progress forward annually through the lifetime of the policy, with the initial conditions at time zero (annuitization) of $\frac{\partial F_0}{\partial S_0} = P$, the policyholder premium, and  $\frac{\partial I_0}{\partial S_0} = 0$, since the first withdrawal occurs at the end of year one. Substituting these recursion values into Equation \ref{VAcf6}, via Equation \ref{VAcf7}, gives the pathwise delta estimator.

\subsection{Conditional LRM VA Liability Estimator}\label{sectVAclrm}
The second alternative Monte Carlo simulation approach for estimating option Greeks proposed earlier in the report was the (conditional) likelihood ratio method (CLRM). We now outline how the CLRM is used for estimating the sensitivities of liabilities arising on the example VA product.

Firstly, we condition on a realization of the Heston variance and CIR interest-rate processes, in the same way as was performed in Section \ref{sect:CLRE}. The only difference here is that we only condition out to the first valuation date at the end of year one, since the VA contract liabilities are path-dependent (recall Example 2 in Section \ref{sectLRMintro}). That is, we calculate
\begin{equation*}
\bar{\sigma}_1 = \sqrt{\frac{a_{3 3}^2}{1}\int_0^1 V_tdt},\hspace{0.15in}
\bar{\xi}_1 = \exp{(\bar{Y}_1)}\hspace{0.1in}\mbox{and}\hspace{0.1in}
\bar{r}_1 = \int_0^1r_tdt
\end{equation*}
where $\bar{Y}_1$ is given by
\begin{equation*}
-\frac{1}{2}a_{3 1}^2\int_0^1 V_tdt -\frac{1}{2}a_{3 2}^2\int_0^1 V_tdt
     + a_{3 1}\int_0^1\sqrt{V_t}dW_t^{I_1} + a_{3 2}\int_0^1\sqrt{V_t}dW_t^{I_2}.
\end{equation*}
The integrals in the above terms can be approximated by simple numerical quadrature. Then the (implied) shock out to year one is then given by:
\begin{equation*}
Z^{\star} = \frac{\log(S_1 / \bar{\xi}_1 S_0) - (\bar{r}_1 - \bar{\sigma}_1^2/2)\times 1}{\bar{\sigma}_1 \sqrt{1}}.
\end{equation*}
Using this implied shock the delta sensitivity can then be determined using
\begin{equation*}
\Delta^{\mbox{LRM}} = \Bigg(  \frac{Z^{\star}}{S_0 \bar{\sigma}_1 \sqrt{1}} \Bigg) \times \mbox{Liability}
= \Bigg(  \frac{Z^{\star}}{S_0 \bar{\sigma}_1 \sqrt{1}} \Bigg) \times \sum_{t=1}^{T} D_t p^{surv}_t \max (I_t - F_t,0)
\end{equation*}
and similarly for the gamma sensitivity. The square-root of one is shown in the formula to help make the approach clear; it comes from the fact the first cashflow occurs at the end of year one. Thus, this term need not equal one if the product paid out semi-annual withdrawals, say, in which case this term would be $\sqrt{0.5}$.

\subsection{VA Liability Gamma Mixed Estimator}
Finally, we apply the pathwise approach to the CLRM estimator for the delta of the VA product liability in order to derive a mixed gamma estimator, in a similar manner as can be done for a European option. This results in
\begin{eqnarray*}
\Gamma^{\mbox{LR-PW}} &=& \frac{\partial}{\partial S_0} \Delta^{\mbox{LRM}}\\
   &=& \frac{\partial}{\partial S_0} \Bigg( \bigg(\frac{Z^{\star}}{S_0 \bar{\sigma}_1 \sqrt{1}} \bigg) \sum_{t=1}^{T} D_t p^{surv}_t \max (I_t - F_t,0) \Bigg)\\
&=&  \Bigg(\frac{Z^{\star}}{S_0 \bar{\sigma}_1 \sqrt{1}} \Bigg) \sum_{t=1}^{T} D_t p^{surv }_t \hspace{0.02in} I(I_t > F_t)\cdot \Big( \frac{\partial I_t}{\partial S_0} - \frac{\partial F_t}{\partial S_0}\Big)\\
  & & -  \Bigg(\frac{Z^{\star}}{S_0^2 \bar{\sigma}_1 \sqrt{1}} \Bigg)  \sum_{t=1}^{T} D_t p^{surv}_t \max (I_t - F_t,0).
\end{eqnarray*}
All the terms in the above formula are already obtained in calculating the pathwise estimator of the liability delta (except the LRM weight which is easily obtained). 

\section{Numerical Comparison of VA Liability Estimators}\label{VAtests}
Having derived the delta and gamma estimators for the example VA product, we now study how the accuracy of the approaches compare. Five test cases will be considered, labelled A-E in Table \ref{tablevaA}, which give different parameter settings for the Heston and CIR processes and different correlations between the normal shocks driving the variance, interest-rate and equity processes. In all these cases the contract has term $T=30$ years and the ratchet term is the first 10 years of the product lifetime. The income drawn each year by the policyholder is $4\%$ of the Guarantee Base and the initial policyholder premium is \textsterling 10,000. For these tests two separate simulation set-ups were created; one for the bump and revalue and another for the pathwise and CLRM estimates. In the bump and revalue set-up 36,000 simulation paths were used. The bump perturbation size was set at 0.5\%. There is a trade-off to be made here as the smaller the perturbation chosen the less bias in the estimate, however reducing the  perturbation size will increase the variance of the estimator - particularly for the gamma. In the CLRM/pathwise set-up 10,000 variance/interest-rate outer paths with 10 equity paths for each of these outer realizations were used. In both frameworks 20 timesteps per year were used to discretize the Heston and CIR processes. By doing this the bump and revalue and CLRM/pathwise approaches took approximately the same amount of computation time to run, giving a fair basis for comparison of the approaches.

\vspace{0.25in}
\begin{table}[ht]
\centering 
\begin{tabular}{| c | c c c | c c c | c c c |} 
\hline 
Case & $\kappa_{\mbox{V}}$ & $\theta_{\mbox{V}}$ & $\epsilon_{\mbox{V}}$ & $\kappa_{\mbox{IR}}$ & $\theta_{\mbox{IR}}$ & $\epsilon_{\mbox{IR}}$ & $\rho_{\mbox{S-V}}$ & $\rho_{\mbox{S-IR}}$ & $\rho_{\mbox{V-IR}}$ \\ [0.5ex] 
\hline 
A & 2 & 0.04 & 0.15 & 0.4 &  0.04 & 0.1 & -0.7 & -0.3 & 0.2 \\ 
B & 1 & 0.04 & 0.3 & 0.4 &  0.04 & 0.1 & -0.7 & -0.3 & 0.2 \\ 
C & 2 & 0.04 & 0.15 & 0.2 &  0.04 & 0.2 & -0.7 & -0.3 & 0.2 \\ 
D & 1 & 0.04 & 0.3 & 0.2 &  0.04 & 0.2 & -0.7 & -0.3 & 0.2 \\ 
E & 1 & 0.04 & 0.3 & 0.2 &  0.04 & 0.2 & -0.9 & -0.3 & 0.2 \\ 
\hline 
\end{tabular}
\caption{Model settings considered in tests. Heston SV parameters denoted by V subscript, CIR parameters by IR subscript. $\mbox{V}_0=\theta_{\mbox{V}}$, $\mbox{r}_0=\theta_{\mbox{IR}}$.} 
\label{tablevaA} 
\end{table}
\begin{table}[h]
\centering 
\begin{tabular}{| c | c c | c c |} 
\hline 
& \hspace{-0.25in}Sim.&\hspace{-0.35in}Set-up 1 (B\&R) & \hspace{-0.25in}Sim.&\hspace{-0.35in}Set-up 2 (PW/LRM) \\
& Liab.(\textsterling) & St.Err & Liab.(\textsterling) & St.Err\\ [0.5ex] 
\hline 
A & 105.57 & 0.68  & 104.65 & 1.18 \\ 
B & 125.19 & 0.98  & 123.68 & 1.78 \\ 
C & 157.40 & 1.02  & 155.39 & 1.81 \\ 
D & 169.11 & 1.27  & 166.86 & 2.31 \\ 
E & 175.39 & 1.52  & 172.26 & 2.87 \\ 
\hline 
\end{tabular}
\caption{\textbf{VA Liability Estimates}: VA example product.} 
\label{tablevaB} 
\end{table}
In table \ref{tablevaB} the estimates of the liability value under the bump and revalue and CLRM/pathwise approaches are given. This table shows that under both simulation set-ups the values of liability are generally consistent. The bump and revalue approach does give estimates with smaller standard errors though. This is due to the conditional simulation framework required by the CLRM approach, which is less efficient than a standard simulation set-up in estimating the product liabilities. However, it is the estimation of the sensitivity of these liabilities to the key market risk drivers that is the difficult challenge facing insurers.

\vspace{0.25in}
\begin{table}[h!]
\centering 
\begin{tabular}{| c | c c | c c | c c |} 
\hline 
 & B\&R & St.Err & PW & St.Err & CLRM & St.Err\\ [0.5ex] 
\hline 
A & -0.00763 & 0.00013  & -0.00734 & 0.00014 & -0.00781 & 0.00032 \\ 
B & -0.00390 & 0.00014  & -0.00362 & 0.00017 & -0.00402 & 0.00040 \\
C & -0.00812 & 0.00016  & -0.00766 & 0.00019 & -0.00814 & 0.00042 \\
D & -0.00384 & 0.00017  & -0.00351 & 0.00021 & -0.00378 & 0.00050 \\
E & -0.00215 & 0.00019  & -0.00176 & 0.00029 & -0.00211 & 0.00060 \\
\hline 
\end{tabular}
\caption{\textbf{Delta estimates}: VA example product.} 
\label{tablevaC} 
\end{table}

\begin{table}[h!]
\centering 
\begin{tabular}{| c | c c | c c | c c |} 
\hline 
 & B\&R & St.Err & CLRM  & St.Err  & PW-LR & St.Err \\ [0.5ex] 
\hline 
A & $4.12$ & $0.55$  & $3.89$ & $0.47$ & $4.42$ & $0.09$ \\ 
B & $3.25$ & $0.53$  & $2.83$ & $0.89$ & $3.17$ & $0.10$ \\
C & $4.25$ & $0.77$  & $4.52$ & $0.70$ & $5.27$ & $0.12$ \\
D & $2.50$ & $0.75$  & $3.45$ & $1.21$ & $3.83$ & $0.13$ \\
E & $3.13$ & $0.70$  & $1.52$ & $3.77$ & $2.67$ & $0.21$ \\
\hline 
\end{tabular}
\caption{\textbf{Gamma estimates}: VA example product. All numbers above should be scaled by a factor of $1\times 10^{-6}$.} 
\label{tablevaD} 
\end{table}

\begin{figure}[h!]
\centering
\includegraphics[width=5.3in]{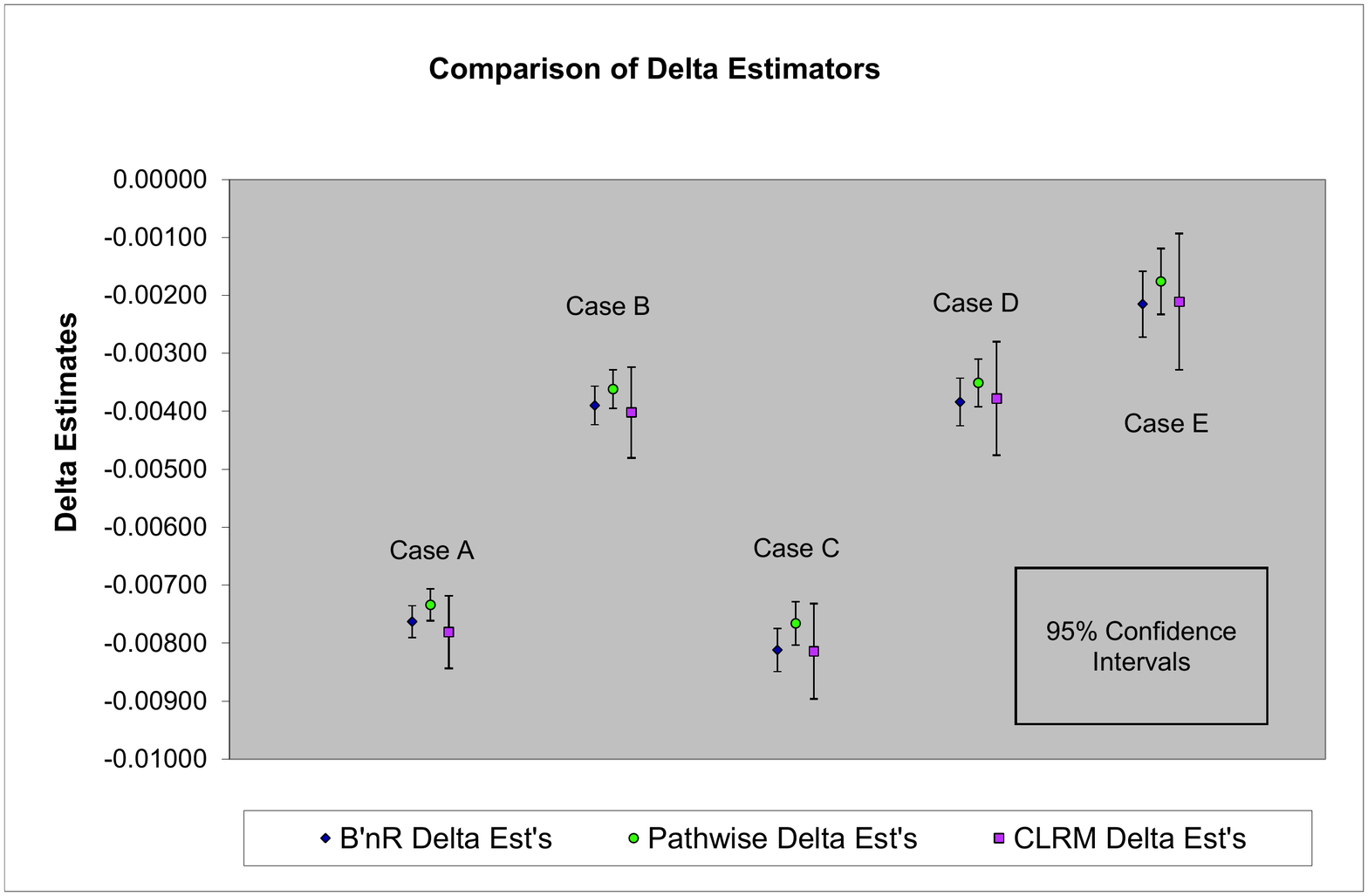}
\caption{}\label{delta 30y plot}
\end{figure}

\vspace{-0.1in}Therefore, let us look at how the approaches perform in estimating the delta sensitivity of the liability in Cases A-E. Table \ref{tablevaC} and Figure \ref{delta 30y plot} give the estimates for the delta sensitivities for each of the approaches discussed earlier. Note that for the bump and revalue estimate, a central difference, using the `bumped up' and `bumped down' simulation paths, rather than a forward difference is used. This can help minimize levels of bias in this estimator. These results show that the bump and revalue and pathwise approaches give similar estimates and standard errors for the delta in each of the given Cases. This is expected, as the pathwise estimator is essentially the small perturbation limit of the bump and and revalue approach. The reason the estimators are not converging to a common value is that they are simulated under the different set-ups: the pathwise results were estimated under the conditional simulation set-up alongside the CLRM estimator, whereas the bump and revalue is from the base run of the bump and revalue set-up. Clearly the CLRM delta estimator is not as efficient as either the bump and revalue or pathwise equivalent (though still giving values consistent with these estimates). This can be explained by the fact that this method does not make use of the payoff function, unlike the pathwise approach. This, however, means the LRM estimator can still be used for options where this is discontinuous (and for second-order sensitivities). The pathwise estimator appears to be as efficient as the bump and revalue, but of course does not require a perturbed simulation run and is unbiased.

\addtolength{\abovecaptionskip}{-8mm}
\addtolength{\belowcaptionskip}{+53mm}
\begin{figure}[h!]
\centering
\includegraphics[width=5.3in]{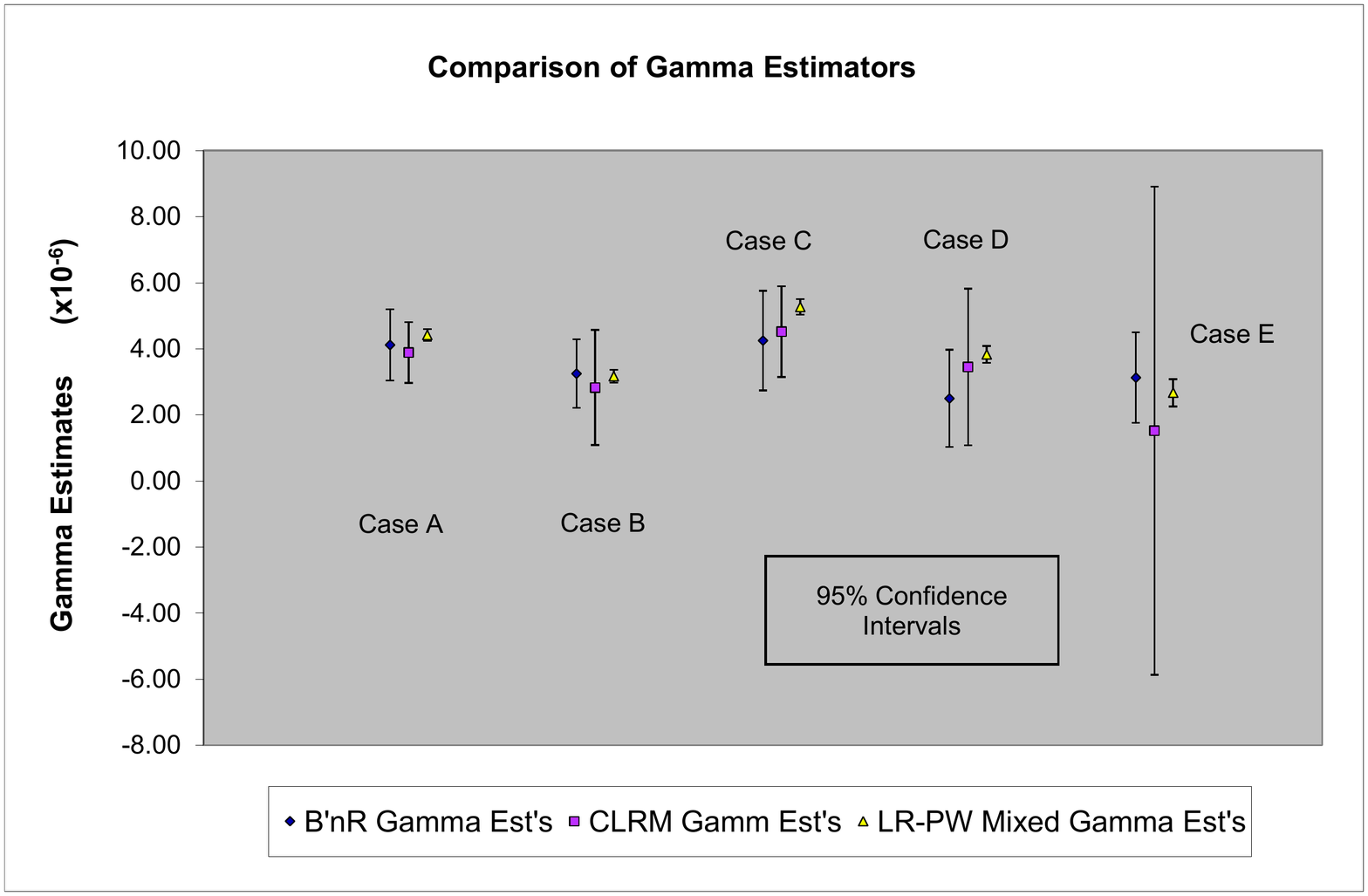}
\caption{}\label{gamma 30y plot}
\end{figure}
\addtolength{\abovecaptionskip}{+8mm}
\addtolength{\belowcaptionskip}{-5mm}

In Table \ref{tablevaD} and Figure \ref{gamma 30y plot} the different estimators for the gamma sensitivity of the VA liability are compared. These results show that both the bump and revalue and CLRM approaches give estimates with fairly large standard errors for cases A-E, with the CLRM being particularly poor for Case E due to the method not being very robust to high levels of $\rho_{\mbox{S-V}}$. The mixed PW-LR estimator gives estimates which are consistent with these two approaches, however they yield a much smaller standard error for each of the Cases considered. 

Thus, by constructing a mixed estimator for the gamma sensitivity, we have managed to gain a significant amount of reduction in variance, compared to the bump and revalue approach. Furthermore, this method is unbiased, hence we do not encounter the problem in choosing the perturbation step-size, where there is a trade-off between trying to reduce bias, yet also keep the size of the standard error of the estimate small. This is consistent with the analysis of the mixed estimators applied to European options in Glasserman \cite{Glass}. 

This significant improvement in efficiency in estimating second-order sensitivities could be of great benefit to practitioners in managing a hedging strategy for their VA books. With the simple bump and revalue approach it is difficult to get an accurate and unbiased estimate of the gamma sensitivity. 
As a result insurers will find it difficult to appreciate how frequently to rebalance delta-hedge positions due to underlying market movements. With a much more accurate and unbiased gamma estimate, the insurer should have a much greater appreciation of the convexity in the liabilities in their VA books and can adapt their hedging strategy portfolio accordingly.

\section{Conclusion} 
With the increasing popularity of VA products and the new Solvency II regulatory framework in Europe, employing an effective hedging strategy for mitigating the risks from marketing such products is a current challenge facing insurers. The recent financial crisis has demonstrated that under turbulent market conditions a hedging portfolio can require much more frequent rebalancing. The widely adopted bump and revalue approach for estimating the Greeks used in such a strategy has some shortcomings, particularly for second-order sensitivities, such as gamma.

In this article some more advanced estimators for VA Greeks have been developed which are unbiased and do not require additional perturbed simulation runs. The mixed estimator developed for the VA Liability gamma sensitivity has much smaller standard errors compared to the bump and revalue method. Furthermore, the bias-variance trade-off in choosing the perturbation size in the bump and revalue is avoided. The estimators are also comparatively more efficient as the required number of Greeks increases.

\newpage
\appendix
\section{Derivation of Stochastic Process for $X_t$.}
In this Appendix, a derivation of the formula for the stochastic process
representing the logarithm of the returns, $X_t$, is given. Recall, that by employing the Cholesky
decomposition the asset price dynamics under the Heston-CIR model were expressed in Section \ref{sect:CLRE} as
\begin{eqnarray*}
\frac{dS_t}{S_t} &=& r_tdt + \sqrt{V_t}dW_t^S\\
     &=& r_tdt + \sqrt{V_t}\bigg( a_{3 1}dW_t^{I_1} + a_{3 2}dW_t^{I_2} + a_{3 3}dW_t^{I_3}\bigg)
\end{eqnarray*}
where $dW_t^{I_1}$, $dW_t^{I_2}$ and $dW_t^{I_3}$ are three independent Brownian motions. 

\vspace{0.1in}
Using this formula the expression given in the article for the stochastic process
representing the logarithm of the returns, $X_t$, can be found as follows. Let $g(t,x)$ be such that $X_t = g(t,S_t) = \ln{S_t}$. Then, 

\begin{equation*}
\frac{\partial g}{\partial t}=0\mbox{, }\hspace{0.3in} \frac{\partial g}{\partial x}=\frac{1}{x}\hspace{0.15in}\mbox{ and } \hspace{0.15in} \frac{\partial^2g}{\partial x^2}=-\frac{1}{x^2}.
\end{equation*}
\vspace{0.2in}

Using It\^{o}'s
formula, $dX_t$, the stochastic process for the logarithm of the asset returns under the Heston-CIR model, is then given by:

\begin{eqnarray*}
\hspace{-1in}dX_t &=& \frac{\partial g}{\partial t}(t,X_t)dt + \frac{\partial g}{\partial x}(t,X_t)dS_t + \frac{1}{2}\frac{\partial^2g}{\partial x^2}(t,X_t)(dS_t)^2\\
     &=& \frac{dS_t}{S_t} + \frac{1}{2}\bigg(-\frac{1}{S_t^2}\bigg)(dS_t)^2\\
     &=& r_tdt + \sqrt{V_t}\bigg( a_{3 1}dW_t^{I_1} + a_{3 2}dW_t^{I_2}
     + a_{3 3}dW_t^{I_3}\bigg)\\
     & & -\frac{1}{2}\frac{1}{S_t^2}\bigg(S_tr_tdt + S_t\sqrt{V_t}\bigg( a_{3 1}dW_t^{I_1} + a_{3 2}dW_t^{I_2} + a_{3 3}dW_t^{I_3}\bigg)\bigg)^2\\
     &=& r_tdt + \sqrt{V_t}a_{3 1}dW_t^{I_1} + \sqrt{V_t}a_{3 2}dW_t^{I_2}
     + \sqrt{V_t}a_{33}\mbox{$dW_t^{I_3}$}\\
     & & -\frac{1}{2}V_ta_{3 1}^2dt -\frac{1}{2}V_t a_{3 2}^2 dt-\frac{1}{2}V_t a_{3 3}^2dt\\
     &=& r_tdt + dY_t -\frac{V_t}{2}a_{3 3}^2 dt + \sqrt{V_t}a_{33}\mbox{$dW_t^{I_3}$} \\
     & & \mbox{with}\\
     \hspace{0.1in} & & \hspace{0.02in}dY_t = -\frac{1}{2}V_t a_{3 1}^2 dt -\frac{1}{2}V_t a_{3 2}^2 dt
     + \sqrt{V_t}a_{3 1}dW_t^{I_1} + \sqrt{V_t}a_{3 2}dW_t^{I_2}.
\end{eqnarray*}


\newpage
\begin{thebibliography}{99}
\bibitem{And}Andersen, Leif B. G., Efficient Simulation of the Heston Stochastic Volatility Model, January, 2007. Available at SSRN: \url{http://ssrn.com/abstract=946405}.
\bibitem{BrGla} Broadie, M., and Glasserman, P., (1996) Estimating security price derivatives using simulation, \emph{Journal of Economic Dynamics and Control} 21:1323-1352.
\bibitem{BroadKaya}Broadie, M. and Kaya, O., Exact Simulation of Option Greeks under Stochastic Volatility and Jump Diffusion Models, \emph{Proceedings of the 2004 Winter Simulation Conference}, The Society for Computer Simulation, 1607-1615.
\bibitem{DuffPan}Duffie, D. and Pan, J., An Overview of Value at Risk, \emph{Journal of Derivatives}, Spring 1997, pp. 7-49.
\bibitem{Glass}Glasserman, P., Monte Carlo Methods in Financial Engineering, 2003. Springer-Verlag, New York.
\bibitem{GlaZh} Glasserman, P., and Zhao, X. (2000) Arbitrage-free discretization of lognormal forward LIBOR and swap rate models, \emph{Finance and Stochastics} 4:35-68.
\bibitem{Glynn} Glynn, P.W. (1987) Likelihood ratio gradient estimation: an overview, pp. 366-374 \emph{Proceedings of the Winter Simulation Conference}, IEEE Press, New York.
\bibitem{Hest}Heston, S., A Closed-Form Solution for Options with Stochastic Volatility with Applications to Bond and Currency Options, \emph{The Review of Financial Studies}, Volume 6 Number 2 pp. 327-343 (1993).
\bibitem{HoCao} Ho, Y.C., and Cao, X.-R. (1983) Optimization and perturbation analysis of queuing networks, \emph{Journal of Optimization Theory and Applications} 40:559-582.
\bibitem{Hobbs}Hobbs, C., Krishnaraj, B., Lin, Y. and Musselman, J., 
Calculation of variable annuity market sensitivites using a pathwise methodology,
\emph{Life \& Pensions}, September 2009.
\bibitem{VApaper}Ledlie, M. C. et.\thinspace al.\thinspace, Variable Annuities, \emph{British Actuarial Journal}, Volume 14 Part 2, 2010.
\bibitem{ReiWeiss} Reimann, M., and Weiss, A. (1989) Sensitivity analysis for simulations via likelihood ratios, \emph{Operations Research} 37:830-844.
\bibitem{SuriZaza} Suri, R., and Zazanis, M. (1988) Perturbation analysis gives strongly consistent sensitivity estimates for the M/G/1 queue, \emph{Management Science} 34:39-64.
\end{thebibliography}
\end{document}